# Characterization of $L2_1$ order in $Co_2FeSi$ thin films on GaAs

**B Jenichen**[1], **T Hentschel**[1], **J Herfort**[1], **X Kong**[1], **A Trampert**[1], **and I Zizak**[2]

[1]Paul-Drude-Institut fuer Festkoerperelektronik, Berlin, Germany
[2]Helmholtz-Zentrum fuer Materialien und Energie, Berlin, Germany

bernd.jenichen@pdi-berlin.

**Abstract**. $Co_2FeSi$/GaAs(110) and $Co_2FeSi$/GaAs(-1-1-1)B hybrid structures were grown by molecular-beam epitaxy (MBE) and characterized by transmission electron microscopy (TEM) and X-ray diffraction (XRD). The films contain inhomogeneous distributions of ordered $L2_1$ and B2 phases. The average stoichiometry could be determined by XRD for calibration of the MBE sources. Diffusion processes lead to inhomogeneities, influencing long-range order. An average $L2_1$ ordering of up to 65% was measured by grazing-incidence XRD. Lateral inhomogeneities of the spatial distribution of long-range order in $Co_2FeSi$ were imaged using dark-field TEM with superlattice reflections and shown to correspond to variations of the Co/Fe ratio.

## 1. Introduction

Device concepts based on the spin rather than the charge of the electron are introduced in the field of spintronics. Heusler alloys can be useful for sources of spin injection as a first step for the fabrication of spintronic devices [1]. The Heusler alloy $Co_2FeSi$ has some outstanding properties: it is a ferromagnetic half-metal with a Curie temperature larger than 1100 K and a magnetic moment of $6\mu_B$. Thus, high spin injection efficiency is expected in $Co_2FeSi$/GaAs hybrid structures. The lattice parameter (0.5658 nm) matches that of GaAs (0.5653 nm), i.e. it can be grown epitaxially on GaAs without the formation of misfit dislocations. In order to obtain a maximum spin injection efficiency fully ordered $Co_2FeSi$ films and perfect ferromagnet/semiconductor (FM/SC) interfaces (IF) are needed. Figure 1 shows a schematic view of the cubic $L2_1$ structure of $Co_2FeSi$ and the corresponding electron diffraction pattern of a $Co_2FeSi$ epitaxial film with the incoming electron beam along the <110> direction. The fundamental reflections form a rectangular pattern and are clearly distinguished by their higher intensity. They are not sensitive to disorder. However, the remaining reflections of lower intensity are superlattice (SL) reflections, which may be used for characterization of the long-range order inside the epitaxial film [2].

For the present work $Co_2FeSi$/GaAs hybrid structures were grown by molecular-beam epitaxy (MBE) and investigated by transmission electron microscopy (TEM) in connection with electron energy loss spectroscopy (EELS) and grazing incidence diffraction (GID) of X-rays in order to characterize the phase stability of the FM/SC IF and the structural properties of the $Co_2FeSi$ film. The average stoichiometry can be calibrated by lattice parameter measurements. Although the films are lattice matched to the substrate, they contain inhomogeneous distributions of ordered $L2_1$ and B2 phases. Average $L2_1$ ordering is measured by GID via comparison of fundamental and SL reflections using an information depth smaller than the film thickness. The as-grown $Co_2FeSi$ films are highly but not fully ordered. In a first approximation the degree of average $L2_1$ ordering is increasing with growth temperature. The lateral distribution of ordering is imaged by dark-field TEM using the corresponding SL reflections.

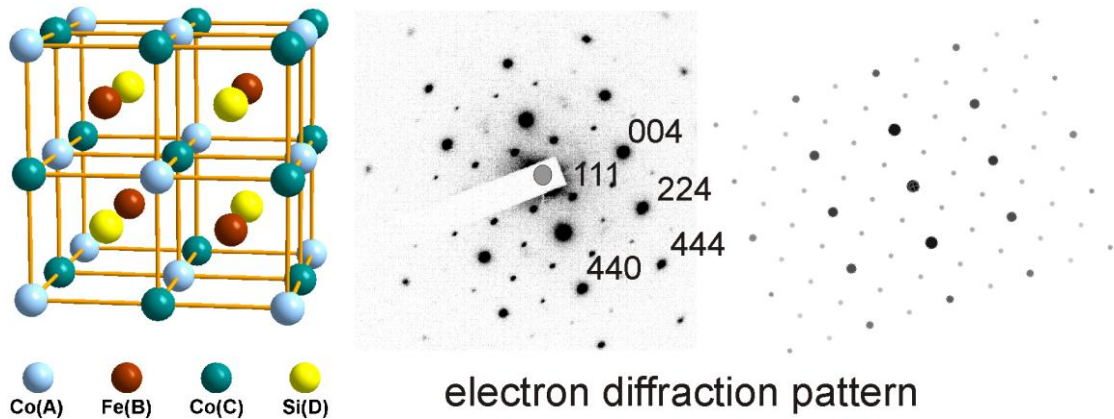

**Figure 1.** Schematic view of the cubic $L2_1$ structure of $Co_2FeSi$ and the corresponding electron diffraction pattern (experiment and calculation using JEMS) of a thin epitaxial $Co_2FeSi$ film grown on a GaAs (110) substrate (incident electron beam directed along <110>).

## 2. Experiment

The $Co_2FeSi$ films were grown on GaAs(110) and on GaAs(-1-1-1)B substrates by MBE [2,3]. The calibration of the fluxes was described in [4]. The growth rate was near 0.1 nm min$^{-1}$. The substrate temperature during MBE growth, $T_S$, was varied between 100 and 350 °C. The nominal $Co_2FeSi$ film thicknesses were 40 nm on GaAs(110) and 15 nm on GaAs(-1-1-1)B. The samples were investigated by dark-field (DF) and high-resolution (HR) TEM. For that purpose cross-sectional TEM specimens were prepared by mechanical lapping and polishing, followed by argon ion milling according to standard techniques. TEM images were acquired with a JEOL 3010 microscope operating at 300 kV. Electron energy loss spectroscopy (EELS) measurements were performed in the TEM with a spot size of 10 nm. The Co/Fe ratio was determined by the analysis of the Fe-$L_{2,3}$ and Co-$L_{2,3}$ edges after background subtraction. Since the analysis was carried out without standards only the lateral changes of the composition ratio were determined in line scans. We determined the average long-range order using a comparison of the integrated intensities of SL and fundamental reflections [5] measured by GID of X-rays. We used the 111, 222, and 220 reflections with synchrotron radiation (energy 6900 eV, wavelength 0.179687 nm) at the beamline KMC2 of the electron storage ring BESSY II of the Helmholtz-Zentrum Berlin.

## 3. Results and discussion

A perfectly ordered $Co_2FeSi$ lattice is needed to obtain the extraordinary properties of this half-metallic Heusler alloy. In order to determine the long-range order by XRD or TEM we can distinguish different types of diffraction peaks: the 220 and the 444 reflections are fundamental (i.e. not sensitive to disorder) whereas the 333, 222 and the 111 reflections are SL reflections [2,5,6]. The 222 reflection arises when at least the CsCl-type B2 order is present in the $Co_2FeSi$ lattice, whereas the 111 or the 333 reflection can be found only in regions of $L2_1$ order. $Co_2FeSi$/GaAs(-1-1-1)B samples are convenient for investigation of SL reflections by TEM, however SL reflections cannot be investigated by GID for this sample orientation. Figure 2 shows the diffraction pattern and corresponding DF micrographs of a nominally 15 nm thick $Co_2FeSi$ film grown on GaAs(-1-1-1)B. The micrograph of the fundamental 444 reflection is more homogenous whereas the 222 and 333 SL reflections exhibit typical inhomogeneities demonstrating laterally disordered regions. Such inhomogeneities were shown by EELS to correspond to local deviations of stoichiometry (Fe/Co ratio) [6].

Dynamical effects prevented quantitative analysis of the DF TEM micrographs. However, in the $Co_2FeSi$/GaAs(110) samples we found the SL reflections well oriented for GID measurements, i.e. their diffracting netplanes are perpendicular to the surface. GID has the advantage of a limited information depth for incidence (and/or exit) angles below the critical angle. In our case, this information depth is smaller than the film thickness of 40 nm, i.e. using GID we measure the region of the film near the surface and exclude an influence of the substrate. Dynamical effects can be avoided in GID. Figure 3 displays the three types of maxima of $\omega/2\Theta$-scans of a $Co_2FeSi$ film grown on GaAs(110) at a substrate temperature $T_S = 250°C$.

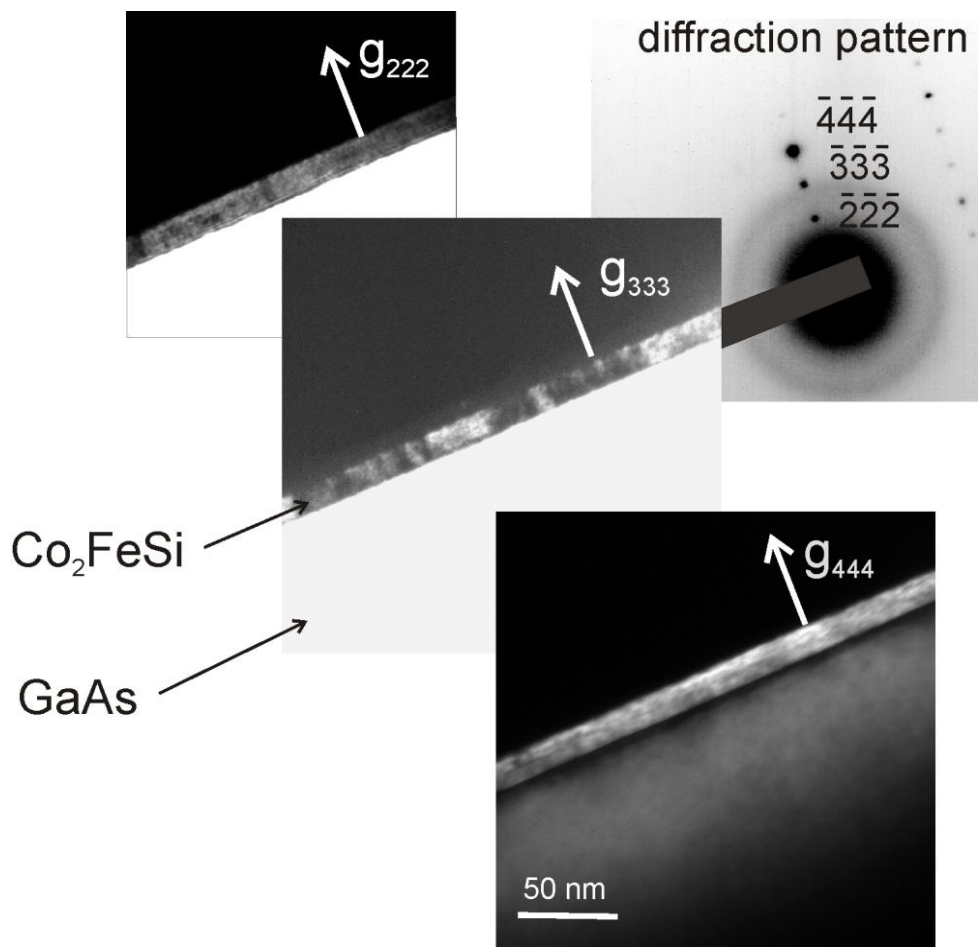

**Figure 2.** Electron diffraction pattern and dark-field TEM micrographs demonstrating the use of the SL maxima 222 and 333 for imaging the lateral distribution of ordered regions in a $Co_2FeSi$ film on GaAs(-1-1-1)B substrate ($T_S = 275$ °C).

On such a basis, we determined the average ordering in three of our samples grown on GaAs(110) at substrate temperatures $T_S = 100$ °C, $T_S = 200$ °C, and $T_S = 250$ °C (see Table 1). The averaging was performed over the whole information depth of the X-rays, which is however smaller than the film thickness, and laterally over an area of several mm$^2$. The region near the FM/SC IF had to be excluded in order to avoid contributions of the GaAs lattice. The result of about 65% L2$_1$ ordering for $T_S$ = 200°C and $T_S$ = 250°C seems reasonable for thin films although for bulk material tempered at high temperatures for long durations, a higher degree of order was reported. Such annealing is not possible for nanometer thick films because of accompanying diffusion processes, which lead to non-stoichiometry near the FM/SC IF. On the other hand, the growth at higher substrate temperatures leads to an increased order in the as-grown film as shown in Table 1 [2, 6]. Often. HRTEM micrographs are taken as evidence for the high structural quality of Heusler alloy films. However in this mode of operation of the TEM many reflections interfere. The SL reflections are of low intensity compared to the fundamental ones. One has to pay attention to minor changes in the interference contrast in order to observe effects due to inter-diffusion and ordering.

**Table 1.** Percentages of average long-range order $S_{B2}$ and $S_{L21}$ of Co$_2$FeSi films grown on a GaAs(110) substrate at different substrate temperatures $T_S$. Regions of B2 and L2$_1$ order may overlap.

| $T_S$ (°C) | $S_{B2}$ (%) | $S_{L21}$ (%) | Error (%) |
|---|---|---|---|
| 100 | 32 | 48 | ±2 |
| 200 | 61 | 65 | ±2 |
| 250 | 38 | 63 | ±2 |

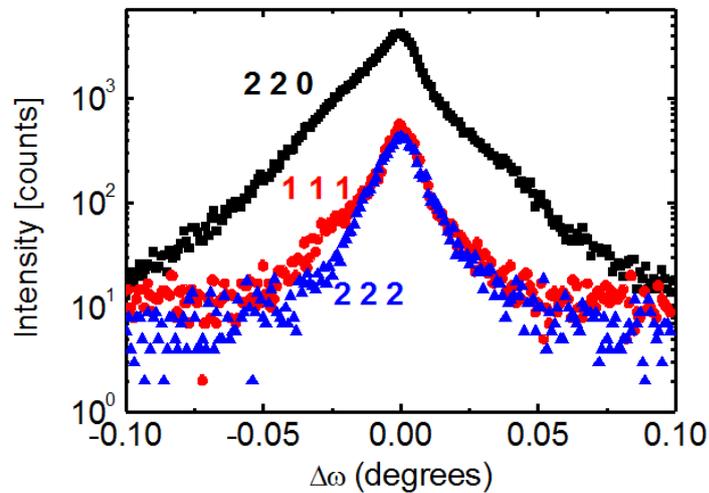

**Figure 3.** Grazing incidence X-ray diffraction peaks of a 40 nm thick Co$_2$FeSi film on GaAs(110) grown at $T_S$ = 250°C. The measurements are performed at an incidence angle below the critical angle of Co$_2$FeSi. Therefore the information depth is below 10 nm. The fundamental 220 reflection (squares) and the 222 (triangles) and 111 (circles) SL reflections are given.